\documentclass[a4paper, 10 pt, conference]{ieeeconf}  
\IEEEoverridecommandlockouts                              

\usepackage[pass,a4paper]{geometry}

\makeatletter
\let\NAT@parse\undefined
\makeatother
\usepackage[square, sort&compress, comma, numbers]{natbib}

\usepackage{graphicx}        
\usepackage{multicol}        
\usepackage[bottom]{footmisc}
\usepackage{tikz}
\usepackage{caption} 
\usepackage{psfrag}
\usepackage{url}
\usepackage{algorithmic}
\usepackage{algorithm}
\usepackage{array}
\usepackage{bm}
\usepackage{xspace}
\usepackage{color}
\usepackage{balance}
\usepackage{amsmath,siunitx,bm,amssymb}
\usepackage{setspace}
\usepackage{framed}
\usepackage{subcaption}
\usepackage{xspace}
\usepackage{tabularx}
\usepackage{multicol,multirow}
\usepackage{booktabs} 
\usepackage[bookmarks=true]{hyperref}
\usepackage[hyphenbreaks]{breakurl} 

\DeclareGraphicsExtensions{.eps} 
\graphicspath{{figures/}{./}}
\newcolumntype{Y}{>{\centering\arraybackslash}X}
\newcolumntype{R}{>{\raggedleft\arraybackslash}X}
\newcommand{\mss}{\si{\metre \per \square \second}}

\newcommand{\p}{\ensuremath{\bm{\theta}}}

\newcommand{\pmin}{\ensuremath{\bm{\theta}_{\text{min}}}\xspace}
\newcommand{\pmax}{\ensuremath{\bm{\theta}_{\text{max}}}\xspace}

\newcommand{\at}{\ensuremath{\tilde{a}\xspace}}

\DeclareMathOperator{\diag}{diag}

\DeclareMathOperator{\capval}{cap}

\title{Effects of Controller Heterogeneity on Autonomous Vehicle Traffic}

\author{Matthew Le Maitre and Amanda Prorok
\thanks{All authors are with the Department of Computer Science and Technology, University of Cambridge, UK: {\tt\small \{msl54, asp45\}@cam.ac.uk}. } %
}

\begin{document}

\maketitle
\thispagestyle{empty}
\pagestyle{empty}

\begin{abstract}
Interactions between road users are both highly non-linear and profoundly complex, and there is no reason to expect that interactions between autonomous vehicles will be any different. Given the recent rapid development of autonomous vehicle technologies, we need to understand how these interactions are likely to present themselves, and what their implications might be. 
This paper looks into the impact of autonomous vehicles with differing controllers, focusing specifically on the effects of changing the mean and heterogeneity of controller parameters on three key performance metrics: \textit{throughput}, \textit{passenger safety} and \textit{comfort}. 
Towards this end, we develop a method for systematically sampling vehicle controllers as a function of parameter heterogeneity. In addition to evaluating the impact of heterogeneity on performance, we quantify the relative impacts of controller input parameters on the output performance metrics by means of sensitivity analyses.
The MovSim traffic simulator was used to simulate a realistic traffic system, whilst recording maximum throughput, as well as lane change frequencies and mean absolute accelerations as proxies for safety and comfort.
Our results reveal that traffic performance is primarily affected by the heterogeneity of vehicle \textit{target velocities}, as well as by the mean values of a very small subset of the parameters, of which the target velocity is by far the most significant.
\end{abstract}


\section{Introduction}
    \label{sec-intro}
    
As different car manufacturers develop their own `in-house' cruise control
algorithms and approach ever higher levels of autonomy, autonomous vehicles will have different corresponding controller behaviours---standardising controller behaviour is likely to be impossible, not least because driving styles (and therefore customer expectations) differ dramatically between individuals, cultures and locations~\cite{awad2018moral}. Interactions between human drivers are complex and highly non-linear; and, in the absence of significant intervention from regulators, there is no reason to expect that autonomous vehicles will be any different. Given the rate of development of autonomous vehicle technologies, we need to understand how these interactions are likely to present themselves, and what their implications will be. 

Analysis of vehicular traffic flow has attracted the attention of statistical and physical research communities since the early 1960s~\cite{drake1966statistical}. For the purpose of simplicity, the bulk of this work traditionally focuses only on \textit{homogeneous} traffic flow, whereby a traffic system is composed of only one type of vehicle (with homogeneous dimensional and dynamic properties)~\cite{hyldmar_Fleet_2019}.
More recent studies on heterogeneous controller interactions look at the relative impacts of a small subset of parameters~\cite{qin2017fundamental, kumar2015short}.
However, there is a dearth of work investigating the impacts over a \textit{comprehensive range of controller parameters}, which is crucial to vehicle controller manufacturers in the process of designing safe and comfortable driving behaviours.

In this paper, we study the impact of autonomous vehicles with differing
controllers sharing road networks. We focus specifically on the effects of
changing the \textit{heterogeneity} of controller parameters on three key performance metrics: \emph{passenger safety, comfort} and \textit{throughput}. 
Towards this end, we develop a method for systematically sampling vehicle controllers as a function of parameter heterogeneity. In addition to evaluating the impact of heterogeneity on performance, we quantify the relative impacts of controller input parameters (and groups of parameters) on the traffic performance metrics by means of sensitivity analyses using the
One-Factor-at-a-Time (OFAT)~\cite{razavi2015we} and Sobol~\cite{sobol2001global} methods.
The MovSim (Multi-model open-source vehicular-traffic Simulator) traffic
simulator was used to simulate a realistic traffic system~\cite{movsimGithub, treiber2013traffic}, whilst recording maximum throughput, as well as lane change frequencies and mean absolute accelerations as proxies of both safety and comfort.
Our results reveal that to maximise traffic safety, as well as increase
throughput by up to $15\%$, controller manufacturers need mainly compare vehicle \textit{target velocities} to minimise heterogeneity along this dimension.


\section{Vehicle Control Model}
    \label{sec-controller}

When simulating traffic systems, a common strategy is to split vehicle
controllers into two components (i.e., subtasks~\cite{toledo2007driving}):
\begin{itemize}
    \item Longitudinal Control for Car Following (CF)
    \item Lateral Control for Lane Changing (LC)
\end{itemize}

The methods used to solve each of these subtasks are of vital significance to traffic flow simulations, and have been studied for several years~\cite{greenshields1934photographic}. Until recently, research into LC was overshadowed by research into CF models, however, this has reversed due to mounting evidence of both the negative impact of LC manoeuvres on traffic safety~\cite{winsum1999lane,pande2006assessment} and its close relation to macroscopic traffic flow characteristics~\cite{ahn2007freeway}.

Two models in particular are used for the controllers in this paper: the \textit{Enhanced Intelligent Driver Model} (E-IDM) for longitudinal, car following behaviour, and the \textit{Minimizing Overall Braking decelerations Induced by Lane changes} (MOBIL) model for lateral, lane changing behaviour. As each of these models are central components of the simulated vehicle controllers, each is covered in turn in the following paragraphs.

\subsection{Longitudinal Control -- E-IDM}

Microscopic `follow-the-leader' paradigms for LC comprise the largest proportion of traffic models~\cite{van2015genealogy}. Initially, these models successfully described only dense traffic, and even then, the gap between vehicles did not necessarily relax to an equilibrium value~\cite{gazis1961nonlinear}. Later developments rectified this using an `optimal velocity function' which incorporated a desired velocity for vanishing interactions~\cite{newell1961nonlinear}, but could lead to extreme accelerations. This was partially rectified by the `Intelligent Driver Model' (IDM), which behaves sensibly at all traffic densities~\cite{treiber2000congested}, but can result in high decelerations after a vehicle cuts in front. This was resolved in the `Enhanced-IDM' model (used in this paper) by implementing a more relaxed reaction to cut-in manoeuvres without losing the mandatory property of being essentially crash free~\cite{Kesting2010EnhancedIDM}.

For an ego-vehicle $\alpha$ at position $x_{\alpha}(t)$ travelling at
    velocity $v_{\alpha}(t)$ with a gap $s_{\alpha}$ to the preceding vehicle,
    which it is approaching at velocity $\Delta v$, the IDM (a precursor to
    E-IDM) describes a vehicle's acceleration using a single continuous function:
    \begin{equation}
        \label{background-eqn-idm}
        \ddot{x}_{\alpha} = \frac{d^2x_{\alpha}}{dt^2} =a\,\left(1-\left({\frac {v_{\alpha }}{v_{0}}}\right)^{\delta }-\left({\frac {s^{*}(v_{\alpha },\Delta v_{\alpha })}{s_{\alpha }}}\right)^{2}\right),
    \end{equation}
    where:
    \begin{equation}
        \label{background-eqn-idm-subin}
        s^{*}(v_{\alpha },\Delta v_{\alpha })=s_{0}+v_{\alpha }\,T+{\frac {v_{\alpha }\,\Delta v_{\alpha }}{2\,{\sqrt {a\,b}}}}.
    \end{equation}

The IDM was developed to guarantee crash free conditions even in the worst possible case, i.e., when the driver of the leading vehicle suddenly brakes with the maximum possible deceleration. In reality, crash-free driving is regularly observed with headway times of significantly less than one second (which IDM would deem critically unsafe)~\cite{treiber2006understanding}. 
E-IDM fixes this inconsistency by adding only one additional parameter, the \textit{coolness}, $c$, to determine the appetite of the ego-vehicle controller for risk, with $c=1$ resulting in no sensitivity to small gaps where the velocity difference is zero (i.e.~too relaxed), and with $c=0$ resulting in pure IDM behaviour (i.e., too conservative). To summarise, the E-IDM parameters are briefly detailed in Table~\ref{background-table-params-eidm}.
    
    \begin{table}[tb]
    {\small
        \centering
        \begin{tabularx}{\columnwidth}{llX}
        \toprule
        \multicolumn{2}{l}{\textbf{Variable}}   & \textbf{Description}
        \\ \midrule $v_{0} $&$\in \mathbb{R}_{>0}  $ & Desired velocity in free traffic \\
        $T $&$\in \mathbb{R}_{>0}$ & Minimum safe headway time between vehicles        \\
        $a $&$\in \mathbb{R}_{>0}$ & Maximum vehicle acceleration \\
        $b  $&$\in \mathbb{R}_{>0}$ & Comfortable vehicle deceleration \\
        $\delta  $&$\in \mathbb{R}_{>0}$ & Functional form of the acceleration profile \\
        $s_{0}  $&$\in \mathbb{R}_{>0} $ & Minimum safe headway distance \\
        $c   $&$\in [0,1)    $ & Determines the appetite for risk \\ 
        \bottomrule
        \end{tabularx}
        \caption{Summary of all parameters in the E-IDM model.}
        \label{background-table-params-eidm}
        }
    \end{table}

\subsection{Lateral Control -- MOBIL}

The more recently discovered impact of LC manoeuvres on traffic safety~\cite{winsum1999lane} and on traffic breakdowns ~\cite{cassidy2005increasing} has generated a surge in LC model research. Kesting et al.~proposed a novel, simplified method of quantifying LC advantages using only single-lane accelerations~\cite{kesting2007general}. The present a decision making model, MOBIL (Minimising Overall Braking Induced by Lane changes), which assesses the possibility, necessity and desirability of a LC. MOBIL is fully acceleration-based, leading to a dramatically simplified (and therefore parsimonious) decision-making process, with highly intuitive parameters. 

    \begin{figure}[tb]
        \centering
        \includegraphics[width=\columnwidth]{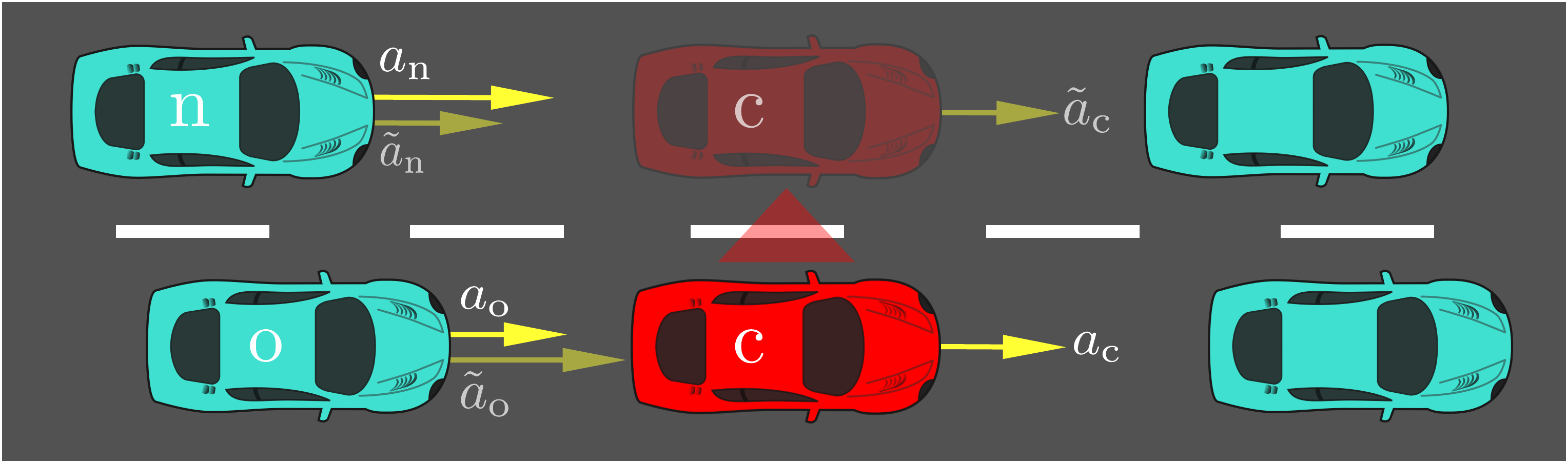}
        \caption{Diagrammatic representation of the variables under consideration when the MOBIL model considers whether a LC should be actioned or not. Accelerations are shown, and faded quantities correspond to theoretical locations and accelerations after the LC.}
        \label{background-fig-mobil}
    \end{figure}
    
To decide if a potential LC should be actioned, MOBIL firstly checks a safety criterion: $\tilde{a}_n \geq -b_{\text{safe}}$, where $\tilde{a}_n$ is the theoretical acceleration of the new upstream follower vehicle after the LC and $b_{\text{safe}}$ is the maximum permissible safe deceleration. 
Next, MOBIL applies one of two fundamental sets of incentive criteria, depending whether lane changing is asymmetric or symmetric (in this paper, asymmetric driving will be assumed). The incentive criterion typically determines whether a LC improves the local traffic situation of the controller and its immediate follower vehicles, as determined by:
\begin{equation}
    \label{background-eqn-mobil-sym}
    [\at_c - a_c] + p\Big( \, [\at_n - a_n] + [\at_o - a_o] \, \Big) > a_{\Delta},
\end{equation}
where $a_c, a_n$ and $a_o$ give the current accelerations of the ego vehicle, new follower, and old follower respectively, and their tilde-capped equivalents, $\at_c$, $\at_n$ and $\at_o$, correspond to the theoretical accelerations after the LC. The $a_{\Delta}$ term prevents LCs if the overall advantage is marginal compared with a `keep in lane' directive. The parameter $p$ is a measure of politeness to other traffic, whereby positive values induce altruism, $p=0$ induces egoism, and negative values induce aggressiveness. A typical LC situation is shown in Figure~\ref{background-fig-mobil}, showing the model parameters in context.

The asymmetric rules are formulated by introducing an additional parameter, a lane bias, $a_{\text{bias}}$. A summary of the parameters required by MOBIL, and a brief description for each, is detailed in Table~\ref{background-table-params-mobil}.

    \begin{table}[tb]
    {\small
        \centering
        \begin{tabularx}{\columnwidth}{llX}
        \toprule
        \multicolumn{2}{l}{\textbf{Variable}}   & \textbf{Description} \\ 
        \midrule 
        $p$ & $\in \mathbb{R}$ & Politeness \\
        $a_{\Delta} $&$\in \mathbb{R}_{>0}     $ & Acceleration gain required for a LC \\
        $a_{\text{bias}}$&$ > a_{\Delta}$ & Const. acceleration bias towards slower lanes \\
        $v_{\text{crit}} $&$\in \mathbb{R}_{>0}$ & Velocity below which traffic is `congested' \\
        $b_{\text{safe}} $&$\in \mathbb{R}_{>0}     $ & Maximum permitted safe deceleration \\ \bottomrule
        \end{tabularx}
        \caption{Summary of all parameters in the MOBIL lane changing model.}
        \label{background-table-params-mobil}
        }
    \end{table}


\section{Method}
    \label{sec-method}

Our method consists of \textit{(i)} a procedure to generate fleets of vehicles with heterogeneous controllers (described in Section~\ref{sec-model}) and \textit{(ii)} two complementary sensitivity analysis techniques (described in Section~\ref{sec-sensitivity}). Our performance metrics are described in~\ref{sec-perf} and the methods used to record performance values are detailed in Section~\ref{sec-setup}.

\subsection{Sampling Heterogeneous Controllers}
    \label{sec-model}
    
To generate heterogeneous controllers, all control parameters were sampled from independent Gaussian distributions with a fractional standard deviation given by the hyperparameter $\sigma$, and a mean given by the hyperparameter $\overline{\p}$ that lies between the hyperparameters $\pmin$ and $\pmax$:
    \begin{equation}
        \p = \capval\Big(\mathcal{N}\left(\overline{\bm{\p}}, \diag\left[\bm{\Sigma}\right]\right)\Big), \label{eqn-sampling}
    \end{equation}
    where $\mathcal{N}$ is a multivariate Gaussian, and where:
    \begin{align}
        \bm{\Sigma} &= \frac{\sigma}{2} \left(\pmax - \pmin\right), \\
        \pmin^{(i)} &\leq \overline{\p}^{(i)} \leq \pmax^{(i)} \qquad \forall i.
    \end{align}
    The $\capval$ function simply caps the values between the minimum and
    maximum allowable values:
    \begin{equation}
        \capval^{(i)}(\p) =
        \begin{cases}
            \pmax^{(i)} & \text{if} \quad \p^{(i)} > \pmax^{(i)}, \\
            \pmin^{(i)} & \text{if} \quad \p^{(i)} < \pmin^{(i)}, \\
            \p^{(i)}    & \text{otherwise}.
        \end{cases}
    \end{equation}

There are four hyperparameters: $\overline{\p}$, $\pmin$, $\pmax$, and $\sigma$. The hyperparameter $\overline{\p}$ can be randomly selected, and reflects the mean hyperparameter settings over all the traffic. The \pmin and \pmax hyperparameters represent hard limits enforced on the model parameters (to prevent non-physical or unrealistic vehicles). These values, shown in Table~\ref{tab-imp-eqn-values}, were chosen based on representative values for current traffic and were fixed throughout all simulations.

The hyperparameter $\sigma$ gives the standard deviation of the parameter distribution such that if $\sigma = \frac{1}{n}$, then the allowable range spans $n$ standard deviations for every parameter. To sample vehicles for a traffic system, values of $\p$ are sampled for consecutive controllers, until enough controllers are generated for the whole fleet. Provided that $\sigma \lessapprox 1$, $\sigma$ is a measure of the heterogeneity of the fleet. This condition was maintained throughput all experiments, so $\sigma$ will be used as a measure of parameter heterogeneity throughout this paper.

    \begin{table}[tb]
    {\small
        \setlength{\extrarowheight}{3pt}
        \centering
        \begin{tabularx}{0.48\columnwidth}[t]{XXRR}
        \toprule
        Param. & Units & Min & Max \\ \midrule
        $p                    $ & $\si{\textit{None}}$ & $0.0  $ & $1.0 $ \\
        $a_{\text{bias}}      $ & $\si{\meter\per\square\second}$ & $0.2\footnotemark[1]$ &  $1.0$ \\
        $s_0                  $ & $\si{\meter}$ & $3.0  $ & $5.0 $ \\
        $v_{\text{crit}}      $ & $\si{mph}$ & $40.0 $ & $65.0$ \\
        $a_{\Delta} $ & $\si{\meter\per\square\second}$ & $0.0  $ & $1.0 $ \\
        $b                    $ & $\si{\meter\per\square\second}$ & $4.0  $ & $7.5 $ \\
        $\delta               $ & \textit{None} & $3.0  $ & $5.0 $ \\ \bottomrule
        \end{tabularx}%
        \hfill
        \begin{tabularx}{0.48\columnwidth}[t]{XXRR}
        \toprule
        Param. & Units & Min & Max \\ \midrule
        \multicolumn{1}{l}{~$\vdots$} & 
        \multicolumn{1}{l}{~$\vdots$} &
        \multicolumn{1}{r}{$\vdots$} &
        \multicolumn{1}{r}{$\vdots$~            \vspace{-1.5mm}
        } \\ 
        $L                    $ & $\si{\meter}$ & $3.0$ & $16.5$ \\
        $v_{\text{max}}       $ & $\si{mph}$ & $70.0$ & $150.0$ \\
        $v_0                  $ & $\si{mph} $ & $40.0$ & $100.0$ \\
        $a                    $ & $\si{\meter\per\square\second}$ & $0.5$&$6.5$\\
        $T                    $ & $\si{\second}$ & 2.0 & 2.0 \\
        $c                    $ & $\textit{None}$& $~0.95 $  & $~0.95$ \\ \bottomrule
        \end{tabularx}
        \caption{Minimum and maximum values of the controller parameters (i.e., caps on the components of the parameter vector $\p$).}
        \label{tab-imp-eqn-values}
        }
    \end{table}

\subsection{Sensitivity Analysis}
\label{sec-sensitivity}

Sensitivity analysis aims to quantify the relative impacts of input parameters on the output performance metrics. Two methods are employed in this paper for sensitivity analysis (one for ease of interpretability, the other for rigour), and are briefly reviewed in the following paragraphs.

\subsubsection{One Factor at a Time (OFAT)}
\label{sec-theory-ofat}

A basic but highly intuitive method to estimate first order dependence (due to only one variable) on the input parameters is the OFAT method. OFAT analysis entails evaluating the function (in our case, any of the traffic performance metrics) at a range of values for one parameter, whilst holding all other parameters constant~\cite{razavi2015we}. If there exists a correlation between the input and the output, the output is sensitive to the input. Gradients can be used to compare relative sensitivities.

OFAT analysis is widely criticised if employed in isolation, as it neglects the joint effects of input variables~\cite{saltelli2010avoid}, and more advanced methods are often more appropriate~\cite{czitrom1999one}. For these reasons, it should only be taken as indicative. Nonetheless, OFAT is both common, and computationally efficient, requiring a minimum of $n+1$ function evaluations for an $n$-dimensional factor space.


\subsubsection{Sobol Indices}

The Sobol method is a more rigorous form of global sensitivity analysis. As well as testing for first-order dependencies, the Sobol method can estimate higher order sensitivities (sensitivities to combinations of variables)~\cite{sobol2001global}. The method is based on decomposition of the output variance into summands of variances of the input parameters in increasing dimensionality. Thus, it determines the contribution of each input parameter and their interactions to the overall model output variance. 

\subsection{Performance Metrics}
    \label{sec-perf}

The three performance metrics evaluated throughout this paper are \textit{comfort}, \textit{safety} and \textit{throughput}.

Comfort is a subjective concept which is difficult to define, and harder still to measure. Nevertheless, it has been shown that comfort decreases monotonically as a function of acceleration magnitude~\cite{thompson1952discussion}, and that most people are `comfortable' at accelerations below $2\mss$~\cite{powell2015passenger}. Higher mean accelerations will also (normally) decrease controller margins of error, and increase the likelihood of crashes. For these reasons, mean acceleration magnitude will be measured as a proxy for comfort and safety.

Traffic safety is strongly negatively correlated with the rate of LC manoeuvres~\cite{winsum1999lane,pande2006assessment}, and frequent LCs also increase mean (lateral) acceleration magnitude, and therefore discomfort. For these reasons, LCs per second (LCPS) will be measured as another proxy for comfort and safety. 
 
Traffic throughput will be measured directly. We note that there is a well established negative correlation between LC rates and road throughput~\cite{cassidy2005increasing}. LCs are critical to the formation and propagation of `stop-and-go' oscillations~\cite{ahn2007freeway}, and high-resolution vehicle trajectories have been used to show that LCs are responsible for changing subtle, localized oscillations into substantial disturbances~\cite{zheng2011freeway}.
    
    
    %


\subsection{Experimental Setup}
    \label{sec-setup}
    
We used the MovSim traffic simulator~\cite{movsimGithub, treiber2013traffic}, with a standard `cloverleaf' road layout, which contains all of the important aspects of highway driving: merges, off-ramps and long motorway stretches. The layout is shown in Figure~\ref{details-cloverleaf}. We ran all simulations for two simulation minutes, updating at 20Hz to emulate an autonomous vehicle controller. We used the same simulations to measure LCPS and mean absolute accelerations (these measurements are independent), with a traffic flow rate of 4,400 vehicles per hour, spread evenly over the network. The simulations used to record maximum throughput were run with an increasing traffic flow rate (up to 12,000 vehicles per hour). We recorded peak density-flow values, giving the max throughout.

    \begin{figure}[t!]
        \centering
        \includegraphics[width=\columnwidth]{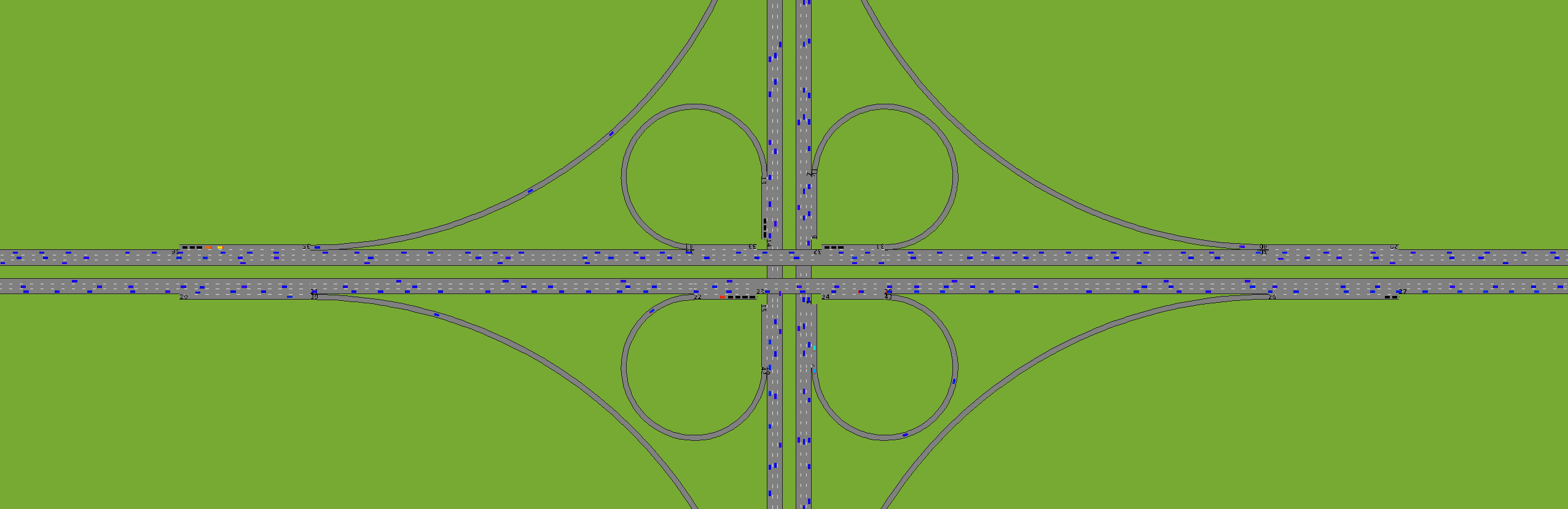}
        \caption{The `cloverleaf' layout of the road network used in simulations.}
        \label{details-cloverleaf}
    \end{figure}
    


To evaluate the effects of heterogeneity, we selected 50 evenly-spaced standard deviations in the range $0 < \sigma < 0.4$, and for each $\sigma$ we ran 10 simulations using values of $\overline{\p}$ sampled according to~\eqref{eqn-sampling}. Each simulation was repeated using 4 different seeds (for a total 200 simulations). During sampling, values of $\overline{\p}$ were kept far from \pmin and \pmax, to ensure that edge effects caused by sampling from a truncated Gaussian are avoided. All three performance metrics were evaluated for every simulation.

To carry out the OFAT analysis, each performance metric was evaluated at 50 different heterogeneities between $0 < \sigma < 0.4$ for each of the 11 model parameters ($b_{\mathrm{safe}}, L, v_0, s_0, \delta, a, v_{\mathrm{crit}}, b, a_{\Delta}, a_{\mathrm{bias}}, b$), giving $550$ simulations. The Pearson Correlation Coefficient (between the output metric and the input heterogeneity $\sigma_i$) is recorded, and a two-tailed Wald hypothesis test is carried out to test if the correlation is significant. The Wald test used a t-distribution of the test statistic, and a null hypothesis of zero correlation. We used a significance level of $p = 0.05$ throughout the paper.

Sobol sensitivity analysis was performed by evaluating each of the 3 performance metrics at 36,000 points in parameter space chosen by using a Sobol Sequence. The results were then used to calculate the Sobol sensitivity indices.


\section{Results}
    \label{sec-results}

In the following, we evaluate the impact of controller heterogeneity on performance, and determine the influence of individual controller parameters.

\subsection{Effect of Controller Heterogeneity}

Figures~\ref{fig-het-vs-lcps}, \ref{fig-het-vs-acc} and \ref{fig-het-vs-throughput}, show the impact of increasing heterogeneity on traffic performance, for LCPS, mean absolute acceleration, and maximum throughput, respectively.

We see that traffic performance drops as heterogeneity increases. The lane change frequency increases, as one would intuitively expect, and correspondingly, the mean absolute acceleration increases up to $0.15\mss$. At an individual simulation level, throughput exhibited a density-flow profile which matched that proposed by Drake~\cite{drake1966statistical}, and the maximum throughput of the system decreased as $\sigma$ was increased, with throughput decreases of over $15\%$ in many cases, and over $10\%$ in all cases.

\subsection{Effect of Individual Controller Parameters}

\subsubsection{One factor at a time (OFAT)}

\begin{figure}[tb]
            \centering
            \includegraphics[width=\columnwidth]{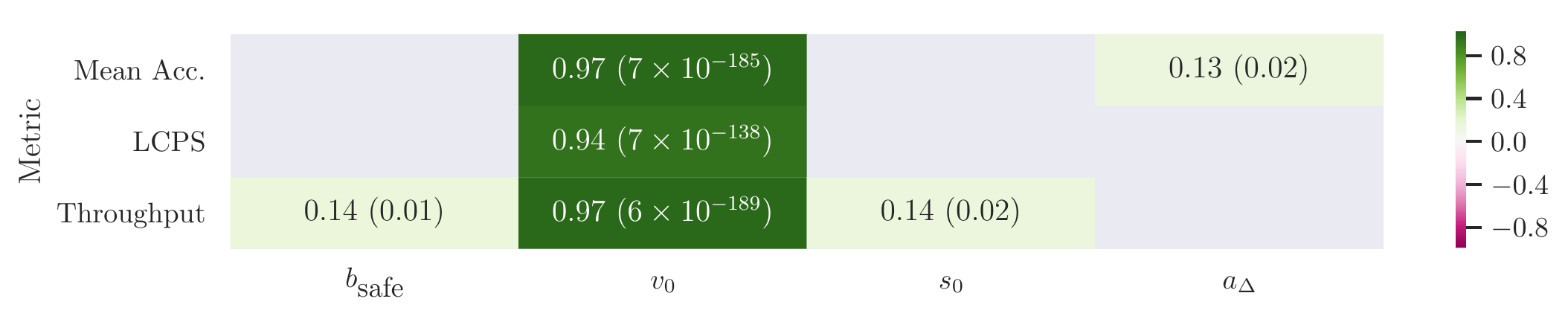}
            \caption{Correlation coefficients and two-sided t-test $p$-values
            (shown in brackets), found using one factor at a time (OFAT)
            analysis. Results are only shown where the $p$-value is less than
            0.05.}
            \label{fig-ofat}
        \end{figure}

The results from the OFAT analysis are shown in Figure~\ref{fig-ofat}. By far the most significant first-order impact on traffic performance is due to the $v_0$ parameter: the target velocity. Much weaker correlations are found for three other parameters, with much lower significance values. As later demonstrated in Section~\ref{results-sobol}, these values are not verified by Sobol analysis, and are hence not significant.

\subsubsection{Sobol analysis}
\label{results-sobol}

\begin{figure*}[tb]
        \centering
        \begin{subfigure}[t]{\textwidth}
            \centering
            \includegraphics[width=\textwidth, trim={0 0 5cm 0},clip]{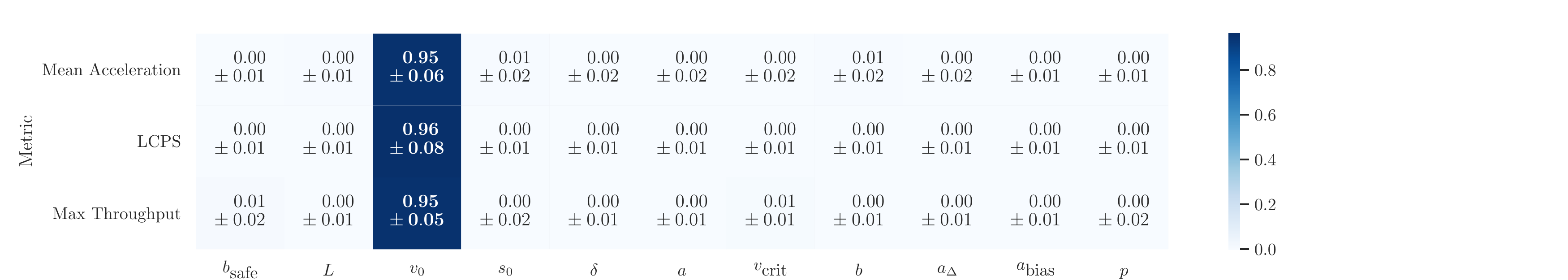}
            \caption{{\footnotesize First order indices for constant means.}}
            \label{fig-sobol-1}
        \end{subfigure}
        \begin{subfigure}[t]{\textwidth}
            \includegraphics[width=\textwidth, trim={0 0 5cm 0},clip]{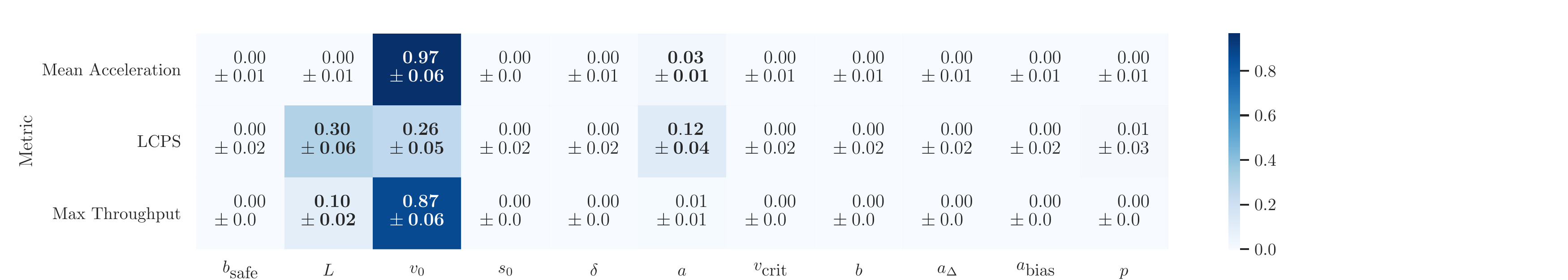}
            \caption{{\footnotesize First order indices for constant heterogeneity.}}
            \label{fig-sobol-mean-v}
        \end{subfigure}
        \caption{(a) First-order Sobol indices for responses of the traffic performance metrics under varying controller heterogeneity. (b) First-order Sobol indices when the mean of each parameter is varied at a constant heterogeneity ($\sigma = 0.2$). We show $95\%$ confidence intervals. The values which have statistically significant nonzero values are given in bold.}
        \label{fig-sobol}
    \end{figure*}

Sobol indices for the first-order effects of each input variable on each of the three traffic performance metrics are shown in Figure~\ref{fig-sobol-1}. As with the OFAT analysis, by far the most significant component comes from the $v_0$ parameter. 
The $95\%$ confidence intervals are consistent with all Sobol indices being zero, except for $v_0$, where the values are all approximately $0.95 \pm 0.06$. 
An analysis of total order Sobol indices (sums of all orders of all indices relevant to each parameter) was also performed, with results similar to the first order results, suggesting that there are no higher order interactions between parameters which impact the performance metrics.

    \begin{figure}[tb]
        \begin{subfigure}[t]{\columnwidth}
        \includegraphics[width=1.00\columnwidth]{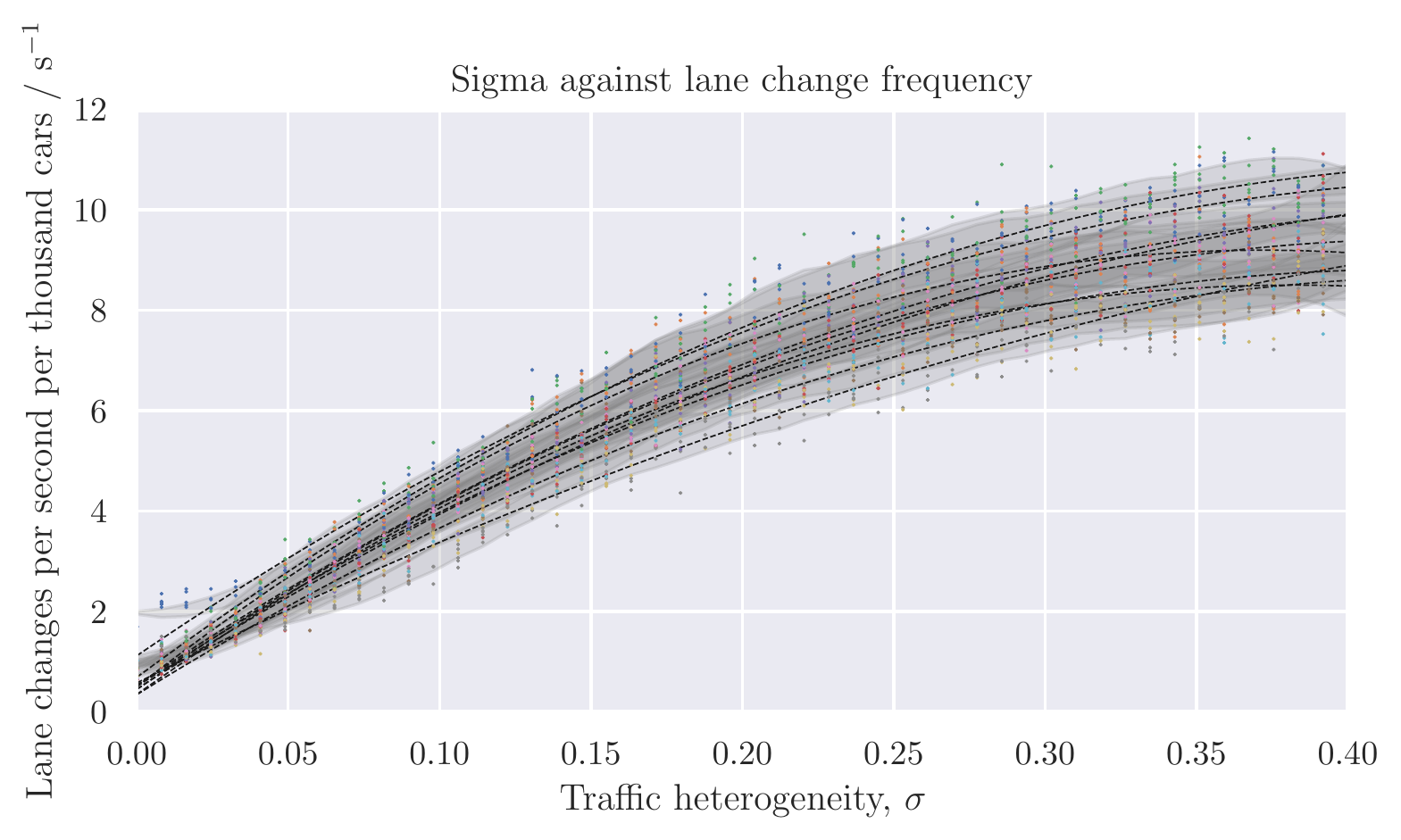}
        \caption{{\footnotesize Lane change frequency against controller parameter
        heterogeneity.}}
        \label{fig-het-vs-lcps}
    \end{subfigure}
    \begin{subfigure}[tb]{\columnwidth}
        \includegraphics[width=1.00\columnwidth]{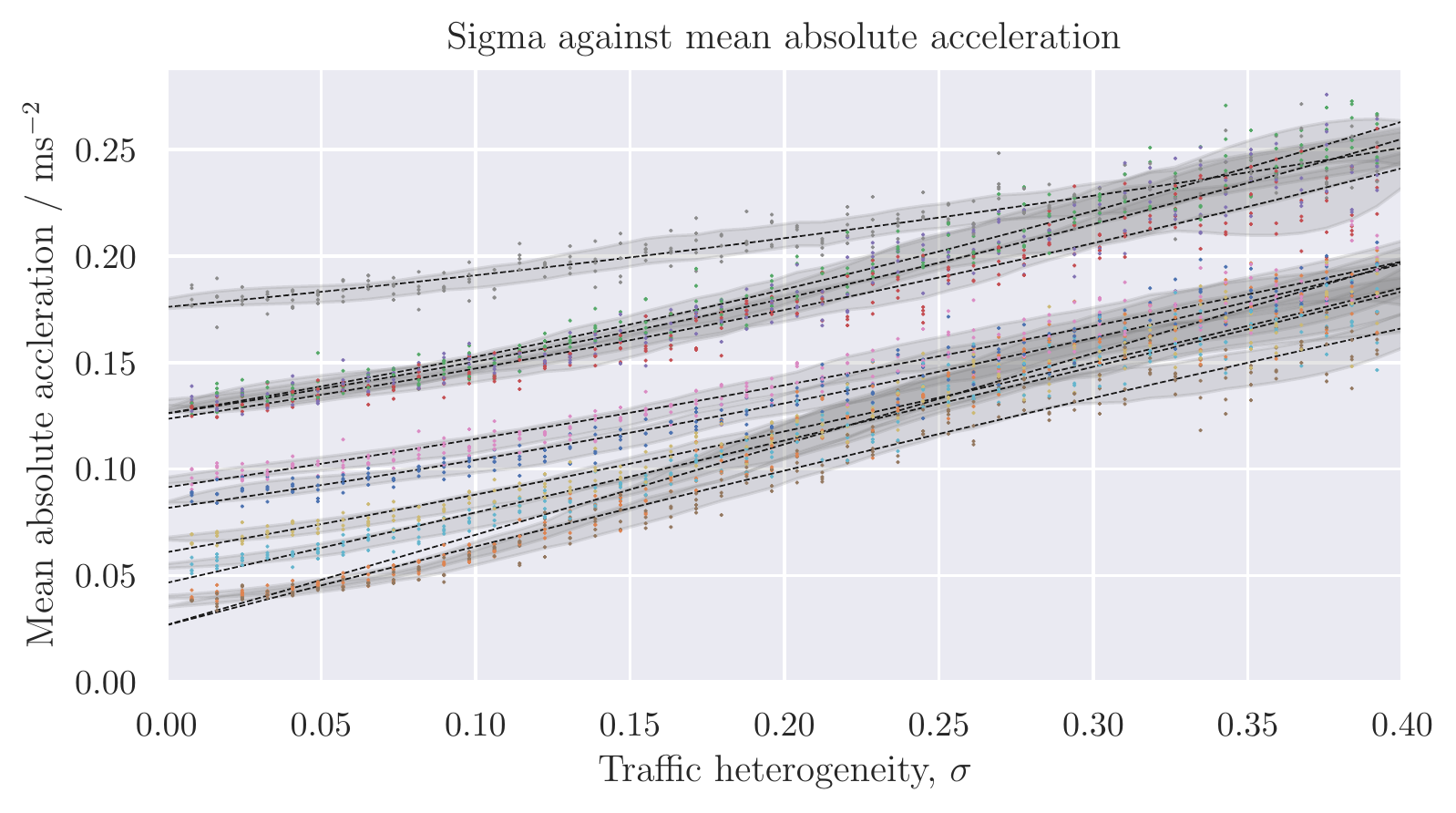}
        \caption{{\footnotesize Mean absolute acceleration against controller parameter
        heterogeneity.}}
        \label{fig-het-vs-acc}
    \end{subfigure}
    \begin{subfigure}[tb]{\columnwidth}
        \includegraphics[width=1.00\columnwidth]{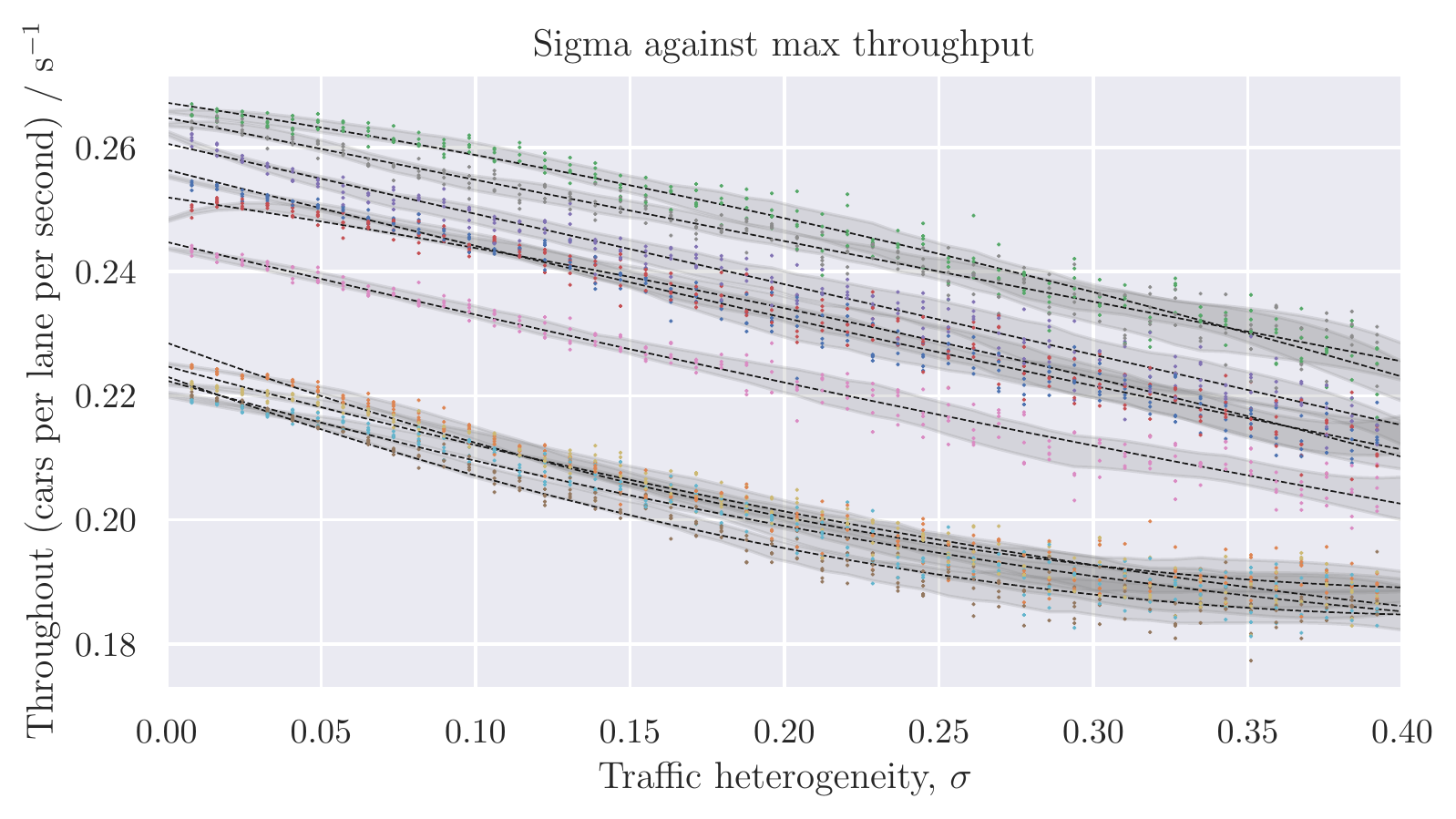}
        \caption{{\footnotesize Maximum vehicle throughput against controller parameter
        heterogeneity.}}
        \label{fig-het-vs-throughput}
    \end{subfigure}
    \caption{Performance metrics as a function of controller parameter
        heterogeneity. Each series corresponds to different mean parameter
        values, and one standard deviation envelopes are shown on all.}
    \end{figure}
        
\subsubsection{Sobol analysis disregarding target velocity}

We also computed the first order and total order Sobol indices \textit{without allowing any variation in the heterogeneity of $v_0$}, to ensure that weaker sensitivities were not being `drowned-out' by the signal from the $v_0$ parameter, and to definitively rule out the non-$v_0$ sensitivities suggested by the OFAT analysis. All indices were zero to within the $95\%$ confidence interval, confirming the results. Second order indices were also computed, and also showed no statistically significant indices.

\begin{figure}[tb]
            \centering
            \includegraphics[width=1.00\columnwidth]{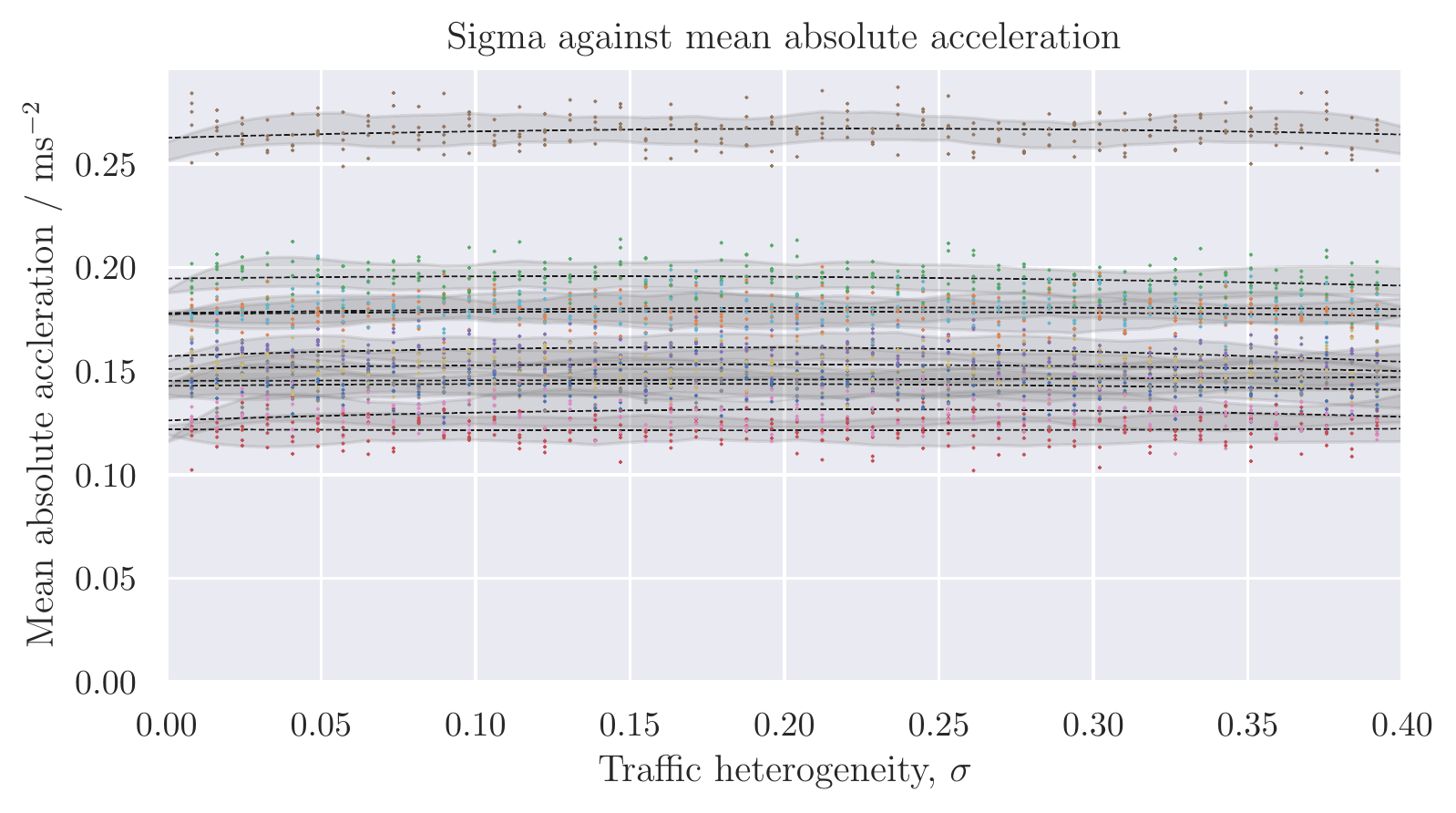}
            \caption{Mean absolute acceleration against controller parameter
            heterogeneity with the exception of fixed heterogeneity in $v_0$,
            from simulated data. Each series corresponds to different mean
            parameter values, and one standard deviation envelopes are shown on
            all.}
            \label{fig-het-vs-acc-nov0}
        \end{figure}

To corroborate this insight, the results in Figure~\ref{fig-het-vs-acc} were re-computed without any variation in $v_0$. Figure~\ref{fig-het-vs-acc-nov0} shows the mean absolute acceleration as a function of varying heterogeneity, and demonstrates no correlation.

\subsubsection{Mean variation under constant heterogeneity}

Figure~\ref{fig-sobol-mean-v} shows first-order Sobol indices when the mean of each parameter is varied at a constant heterogeneity of $\sigma=0.2$. The only parameters that have any effect are length ($L$), target velocity ($v_0$), and vehicle acceleration ($a$). The values that have statistically significant nonzero values are highlighted. Mean absolute acceleration is affected by the maximum acceleration $a$ (as one would expect) and by $v_0$ (possibly because higher velocities require more time spent accelerating). Lane change frequency is only affected by $v_0$, $a$ and vehicle length $L$. Increasing vehicle length reduces LCPS, as rarer, larger gaps are required for lane changes of larger vehicles. The parameter $v_0$ also affects LCPS because faster vehicles change lanes in less time, and $a$ affects LCPS because higher accelerations mean that a higher proportion of LC manoeuvres are both feasible and safe. Maximum throughput is impacted by target velocity $v_0$ and length $L$, as they impact the number of vehicles on the road and their speed.


\section{Discussion}
    \label{sec-discussion}

This work analysed the impact of controller heterogeneity on traffic throughput, mean absolute acceleration and lane changing frequency, the latter two being proxies for both comfort and safety. To this end, we developed a controller sampling method, and applied sensitivity analyses using OFAT and the Sobol method, to determine which subset of parameters was most significant in determining traffic performance, in terms of parameter heterogeneity across all vehicles.

Our results showed that increasing the heterogeneity of controller parameters decreased traffic performance. In particular, we showed that reductions in controller heterogeneity increased maximum road throughput by over $15\%$, and significantly reduced lane change frequencies and mean absolute accelerations, thereby increasing passenger safety and comfort.

Sensitivity analyses showed that the target velocity was by far the most significant parameter, routinely producing Sobol indices over $20$ times as large as the next largest index. For all practical intents and purposes, the target velocity is the only significant parameter pertinent to traffic performance.

Although the considered parameters are specific to the MOBIL/E-IDM based controllers used in this work, it seems unlikely that controller behaviours will differ drastically from the behaviours observed in these simulations. Therefore, any vehicle controllers should be capable of being emulated to a certain extent by this MOBIL/IDM model. If so, there could be significant
repercussions for the autonomous vehicle industry. By cooperating to ensure there is low heterogeneity between target velocities, controller manufacturers should be able to improve comfort, throughput and safety whilst maintaining corporate confidentiality over the finer details of their controllers, as it is unlikely that any of the other parameters will have any significant impact on traffic performance. 




\section*{Acknowledgements}
This work was supported by the Engineering and Physical Sciences Research Council (grant EP/S015493/1). 



\footnotesize{
\bibliographystyle{abbrv}
\bibliography{bib}
}

\end{document}